\documentclass[twocolumn]{webofc}

\usepackage{epsfig}
\usepackage[varg]{txfonts}   
\usepackage{hyperref}
\usepackage{url}
\hypersetup{colorlinks=true,citecolor=blue,urlcolor=blue,linkcolor=blue}
\usepackage{colordvi}\usepackage[usenames]{color}\usepackage{pict2e}
\newcommand{\Put}[3]{\put(#1,#2){\makebox(0,0){#3}}}

\newcommand{\Fbox}[1]{\setlength{\fboxsep}{0.2\fboxsep}\fbox{#1}\setlength{\fboxsep}{5\fboxsep}}
\newcommand{\point}[2]{\psfig{file=point#1,height=#2}}
\newcommand{\PATH}{}
\newcommand{\etal}{{\it et al.}}
\begin{document}
\title{Scattering lengths beyond the nuclear scale and the Efimov effect}

\author{\firstname{F.~Miguel} \lastname{Marqu\'es}\inst{1}\fnsep\thanks{\email{marques@lpccaen.in2p3.fr}}}

\institute{LPC Caen, ENSICAEN, CNRS/IN2P3, Universit\'e de Caen, Normandie Universit\'e, 14050 Caen, France}

\abstract{The interaction of neutrons and nuclei at low energies may potentially lead to scattering lengths several orders of magnitude larger than the effective range of the interaction, well beyond the nuclear scale. If such cases existed, they could lead to the observation of the Efimov effect in nuclei, a remarkable universal phenomenon that has been observed only in atoms.
 The interaction parameters of neutrons scattering off unstable nuclei can be explored in neutron-nucleus systems created after the fast removal of a few nucleons from a slightly heavier beam. The case of the $^{17}$B-$n$ system is considered, and the implications of its potentially huge scattering length on the structure of $^{19}$B as a $^{17}$B-$n$-$n$ Efimov trimer are discussed.}
\maketitle
\section{The Efimov effect}

 Quantum systems, in particular those built by the nuclear force, exhibit intricate structures due to the convoluted nature of the interaction between their constituents. The incomplete knowledge of a force that acts within a range comparable to the sizes of the many constituents, and of the whole many-body system, leads to effective descriptions that rely on a large number of parameters.
 However, some quantum systems are close to the emission threshold of one of the particles, which becomes weakly bound and delocalized. In those cases, the main part of the wave function `tunnels' beyond the interaction range into a region without potential and becomes thus extremely simple, dependent only on the binding energy. Those systems are said to be universal \cite{Hammer2006}, in the sense that different systems governed by very different short-range interactions, provided they lead to the same few parameters in the asymptotic region, can exhibit the same universal states.
 
 In 1970 Efimov found a counterintuitive effect, also universal, that could arise in three-body systems built from such two-body systems near threshold \cite{Efimov1970}.
 Two-body scattering through a potential of range $R$ at momentum $k=\sqrt{2\mu E}$ ($\mu$ is the reduced mass of the system and $E$ the relative energy) is dominated by partial waves with angular momentum $\ell\lesssim kR$. At sufficiently low energies, the only significant contribution comes from $\ell=0$, inducing a phase shift $\delta_0$ in the asymptotic wave function \cite{Scattering}:
\begin{equation}
 \varphi_k(r)\ =\ \sin(kr+\delta_0)/kr	\label{e:delta}
\end{equation}
 The function $\delta_0$ may be expanded at low energy in even powers of $k$ within the effective-range approximation \cite{Scattering}:
\begin{equation}
 k\,\cot\delta_0\ =\ -1/a_s\ +\ r_e\,k^2/2\ +\ \mathcal{O}(k^4)	\label{e:kcot}
\end{equation}
 The low-energy scattering cross-section reads:
\begin{equation}
 \sigma(k)\ =\ 4\pi\,\sin^2(\delta_0)\,/\,k^2	\label{e:scat}
\end{equation}
 giving a length scale to the strength of the interaction in the first term of the expansion, since $\sigma(0)=4\pi a_s^2$.
 
 Therefore, the two-body interaction is represented by the phase shift $\delta_0(k)$ it induces, which is characterized by the two leading parameters of Eq.~(\ref{e:kcot}), the scattering length $a_s$ (a measure of the interaction strength towards zero energy) and the effective range $r_e$ (a measure of the interaction range).
 Negative values of $a_s$ correspond to unbound states of the system for attractive potentials, while positive values correspond to bound/unbound states for attractive/repulsive potentials \cite{Scattering}. For nucleons scattering off light nuclei both parameters are typically of the order of several fm, and thus $|a_s|\sim r_e$.
 
\begin{figure*}
 \psfig{file=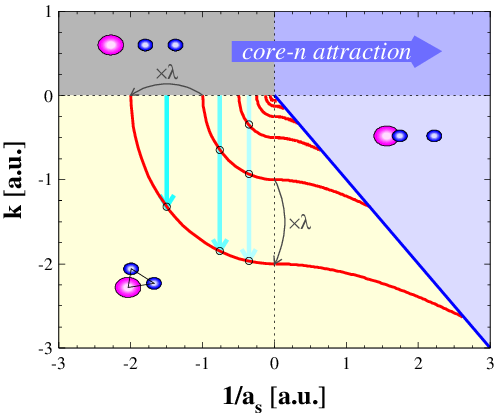,width=\columnwidth} \hfill
 \psfig{file=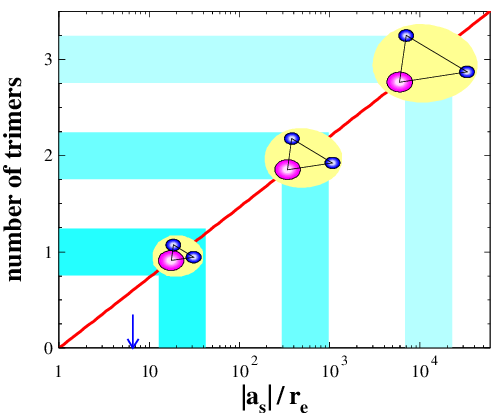,width=\columnwidth}
 \caption{Left: schematic view of the so-called ``Efimov plot'' for a system of a core nucleus plus two neutrons, with its momentum $k$ versus the inverse scattering length $a_s$ of the core-$n$ subsytem; the yellow area represents the region of bound trimers, and the gray and blue areas represent respectively the regions of three-body and dimer-$n$ scattering continuum; the red lines correspond to the trajectory of the first six Efimov trimers (related to each other by an arbitrary $\lambda=2$), and the blue line to the one of the core-$n$ dimer; the cyan arrows represent $a_s$ values that would lead to 1, 2 and 3 trimers. Right: the approximate number of trimers $N\approx\frac{1}{\pi}\ln(|a_s|/r_e)$ as a funtion of the $|a_s|/r_e$ ratio for the core-$n$ dimer (with also $\lambda=2$ in the trimer pictures); the blue arrow indicates the value for the $n$-$n$ system.} \label{f:Ooo}
\end{figure*}
 
 Efimov, on the other hand, considered systems in which the strength of the interaction would overwhelm its range ($|a_s|\gg r_e$), which he called ``resonant''. While trying to build three-body systems from these resonant interactions, he found something unexpected when their size (the hyper-radius, defined as ${\cal R}^2=r_{12}^2+r_{13}^2+r_{23}^2$) was comprised within these two length scales \cite{Hammer2006}:
\begin{equation}
 r_e\ \ll\ {\cal R}\ \ll\ |a_s|	\label{e:hyperR}
\end{equation}
 In the Schr\"odinger equation, several terms with $r_e/a_s$ and ${\cal R}/a_s$ could be then neglected, and the remaining equation was left with an effective three-body attraction of the form $-1/{\cal R}^2$. Therefore, short-range two-body interactions generated a long-range three-body one, and in the case of unbound systems ($a_s<0$) this attraction would lead to ``Borromean'' binding \cite{Pascal2017}.
 Although his prediction was for the simplest case of three identical bosons \cite{Efimov1970}, he soon extended it to other trimers, provided there were at least two resonant interactions out of the three \cite{Efimov1973}. 

 But the most remarkable finding was that the system became scale invariant, since a potential $-1/{\cal R}^2$ scales as the kinetic energy $\mbox{d}^2/\mbox{d}{\cal R}^2$ under a scaling transformation ${\cal R}\rightarrow\lambda{\cal R}$. In fact, if there is a solution at a given $E<0$, {\em any} such transformation would give another solution at energy $E/\lambda^2<0$, and the system would collapse.
 In order to avoid it, Efimov added a boundary condition at ${\cal R}\sim r_e$ through a ``three-body parameter'', which fixed the three-body observables while absorbing the effects of the two-body (and possibly three-body) interactions at short distance \cite{Pascal2017}. Surprisingly, this parameter broke the scale invariance for {\em arbitrary} $\lambda$, but the equation was still invariant under a {\em discrete} set of scale transformations ${\cal R}\rightarrow\lambda^n{\cal R}$, with scaling factors that were integer powers of a fixed $\lambda$ (about $e^\pi\approx22$ for identical bosons \cite{Efimov1970}).

 Therefore, there would be a geometric series of bound states on top of the ground state, with increasing sizes and with energies accumulating at zero:
\begin{eqnarray}
 \psi_{n+1}(\lambda r)	& = & \psi_n(r)		\label{e:psi}	\\
 E_{n+1}				& = & E_n/\lambda^2	\label{e:ene}
\end{eqnarray}
 On the left of Fig.~\ref{f:Ooo} we show the trajectory of the first states of the geometrical series for a core-$n$-$n$ system, in arbitrary units and with $\lambda=2$ for better visibility (since $k\propto\sqrt{E}$ it also scales with $\lambda$).
 This plot illustrates what Efimov called ``discrete scale invariance''. For example, from any point in the trajectory of the first state we can obtain the coordinates of the equivalent point in the trajectory of the $n$th state as $(k,1/a)_n=(k,1/a)_1/\lambda^n$.

 For a pair of particles interacting trough an attractive potential, the X axis of the plot shows increasing attraction. When the dimer is unbound, $a_s$ is negative and starts being small; around the origin of the plot, it goes to $-\infty$, and then back from $+\infty$ to bind the dimer, that will be more bound the smaller $a_s$ (blue line).
 Therefore, the spectacular effect Eimov predicted would happen around this exceptional point at $(0,0)$. For a given effective range, the number of trimers in the series would be larger the bigger $|a_s|$, and at the unitary limit, where $|a_s|\rightarrow\infty$, there would be an infinite number of bound states.

 It is important to note that this phenomenon is independent of the nature or mass of the particles, or of the details and shape of the dimer interaction. The effect is universal. Once the interaction range is given, it depends only on the dimer scattering length and the trimer size through Eq.~(\ref{e:hyperR}). The nature and mass of the particles play a role on the {\em value} of $\lambda$ in Eqs.~(\ref{e:psi},\ref{e:ene}), but not on the occurrence of the effect itself \cite{Efimov1973}.

\section{Atomic observation and nuclear candidates}

 The original Efimov paper introduced the effect for ``{\it the simplest case, three spinless neutral particles of equal mass}'', but then discussed two nuclear candidates, the Hoyle state in $^{12}$C and the triton \cite{Efimov1970}. In the former, the effect would be hindered by the Coulomb repulsion, and in the latter by the small triplet $p$-$n$ scattering length $a_t\approx5$~fm, leading to the conclusion that they would ``{\it feature one such level at best}''.
 The cyan arrows in the Efimov plot of Fig.~\ref{f:Ooo} show particular cases in which only a few trimers would exist. In fact, the discrete scale invariance leads to an upper limit according to $N\approx\frac{1}{\pi}\ln(|a_s|/r_e)$ \cite{Efimov1970}, which is shown on the right of the figure.

 The only known favorable case in nuclear physics would have been the trineutron, with three identical neutral particles and a large $a_s\approx-20$~fm. However, its $|a_s|/r_e\approx7$ (blue arrow in Fig.~\ref{f:Ooo}) would not be enough even for the existence of the first trimer, but more importantly, since neutrons are identical fermions, Pauli repulsion would have suppressed the three-body attraction anyway \cite{Endo2025}. This is the main reason why all exact calculations systematically predict the non-existence of the trineutron, neither as a bound nor as a resonant state \cite{FMM2021}. 
 
 But since Efimov's prediction was universal, the search for experimental confirmation could be undertaken in another field, atomic physics. Although the natural scattering lengths in diatomic systems are usually not large enough to expect the formation of Efimov states, in ultra-cold atom experiments it is possible to modify the scattering lengths over several orders of magnitude by applying a magnetic field thanks to Feshbach resonances \cite{Feshbach}. This allowed to probe the Efimov plot of Fig.~\ref{f:Ooo} not only at a single fixed arrow as in nuclear physics, but over a whole range of scattering lengths.

 The first Efimov trimer was thus observed in 2006 within an ultra-cold gas of Cs atoms \cite{Efimov_CsCsCs}, although the discrete scale invariance remained to be proven. In 2009 a second trimer was identified, allowing to measure a scaling factor $\lambda=21.0(1.3)$ \cite{Efimov_x2}, close to the original prediction of 22.7 for three identical bosons.
 However, a third trimer that would confirm the scaling found between the first two was still elusive. As the energy scales with $\lambda^2$, each additional trimer would be $\approx500$ times less bound than the previous one, and therefore extremely sensitive to temperature variations.
 
 With particles of different masses the scaling factor can be smaller than 22.7, and thus the third trimer easier to observe. However, experiments with heteronuclear mixtures are considerably more challenging.
 One Efimov trimer was observed in a gas of $^{41}$K and $^{87}$Rb atoms by tuning the K-Rb scattering length, both for the K-Rb-Rb (light-heavy-heavy) and K-K-Rb (light-light-heavy) configurations \cite{Efimov_RbKK}.
 Efimov {\em physics} (only one trimer does not test discrete scale invariance) were thus observed for the first time in systems with different masses and only two resonant interactions.
 The Efimov effect was finally confirmed with the observation of the third trimer in a mixture of $^6$Li and $^{133}$Cs atoms, with three Li-Cs-Cs resonances following a geometric series for $\lambda=4.9(4)$ \cite{Efimov_x3a,Efimov_x3b}, in agreement with the predicted value of 4.88.
 
 As to new nuclear candidates, if one avoids Coulomb forces and excludes the trineutron, the only potential candidate would be a system of a core nucleus plus two neutrons, like the one pictured in Fig.~\ref{f:Ooo}. However, the light-light-heavy configuration is the less likely to occur, with a factor $\lambda$ that can reach the thousands\footnote
{A core-core-$n$ system would be much more favorable in terms of masses, with two heavy cores that would exchange the lighter particle, a neutron, like in the Li-Cs-Cs atomic example. However, the effect would be strongly hindered by Coulomb repulsion.} \cite{Pascal2017}.
 Therefore, in order to find a candidate with at least two trimers, Eq.~(\ref{e:hyperR}) would require a core-$n$ system with a scattering length significantly beyond the nuclear scale, since with $\lambda\sim1000$ the size of the second core-$n$-$n$ trimer would be of the order of several thousand fm (${\cal R}_{n+1}=\lambda{\cal R}_n$).
 
\begin{figure}[t]
 \psfig{file=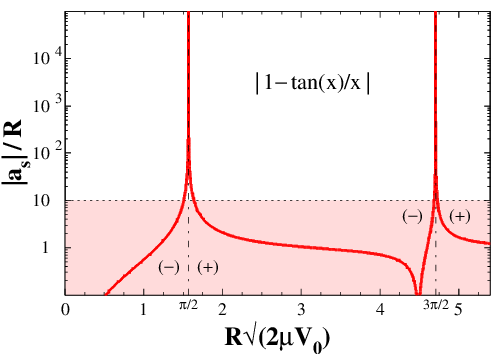,width=\columnwidth}
 \caption{The absolute value of $a_s$ for a particle scattering off a square-well potential of radius $R$ and depth $V_0$ (in units of $R$) as a function of $x=R\!\sqrt{2\mu V_0}$. The scattering length $a_s$ is negative for $x<\pi/2$ and changes sign (in parentheses) subsequently, in particular when it diverges towards $\pm\infty$ at odd multiples of $\pi/2$.} \label{f:a_R}
\end{figure}

 The neutron-neutron scattering parameters $(a_s,r_e)=(-18.5,2.8)$~fm are an exception within the known neutron-nucleus ones, all with scattering lengths and effective ranges of the order of a few fm. 
 But a simple exercise shows that we should not exclude the possibility of finding very large neutron-nucleus scattering lengths. The one of a particle scattering off a square-well potential of radius $R$ and depth $V_0$, as a function of $x=R\!\sqrt{2\mu V_0}$, is given by $a_s=R(1-\tan x/x)$.
 In Fig.~\ref{f:a_R} we have plotted the dimensionless quantity $|a_s|/R$ in logarithmic scale in order to show more explicitly how, for some very specific values of the width and depth of the potential leading to odd multiples of $\pi/2$ in $x$, the scattering length can in practice diverge towards $\pm\infty$.
 But does such a neutron-nucleus system exist in Nature?

\section{Nuclear observables}

 Neutrons produced in some nuclear reactions can be collimated and sent onto stable targets, and their scattering measured at different energies and angles in order to determine the phase shift $\delta_0(E)$, or the leading parameters of its expansion $(a_s,r_e)$.
 But in the search for possible candidates with huge scattering lengths it would be desirable to extend this technique to exotic, unstable nuclei, for which the technique must adopt a different perspective. Short-lived nuclei, as well as the neutron itself, cannot be used as targets, and therefore beam-target scattering experiments are no longer possible. The closest possible scenario is the creation of the neutron and nucleus of choice in the final state of a reaction with a range of low relative energies.
 
 If we assume an initial state in which the neutron and the nucleus 
 are bound within a projectile of similar size, with only a few extra nucleons, and we remove the latter using a high-energy fast removal reaction, we can use the sudden approximation and calculate the scattering amplitude of the final neutron-nucleus system as the overlap integral, leading to the cross-section:
\begin{equation}
 \sigma(k)\ \approx\ k\,\left|\,\int_0^\infty\!\!\psi_i(r)\ \varphi_k^*(r)\,r^2dr\,\right|^2 \label{e:sigmaO}
\end{equation}
 Using the asymptotic scattering wave function for $\varphi_k(r)$ in Eq.~(\ref{e:delta}), the Schr\"odinger equation can be solved to determine the initial wave function of the neutron $\psi_i(r)$ in the projectile, and then the cross-section (\ref{e:sigmaO}) evaluated.
 
\begin{figure*}
 \psfig{file=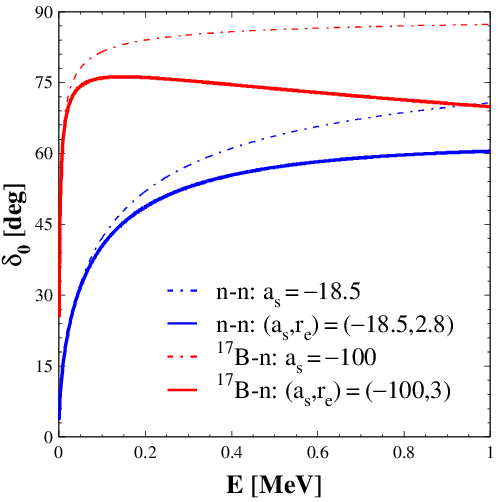,width=\columnwidth} \hfill
 \psfig{file=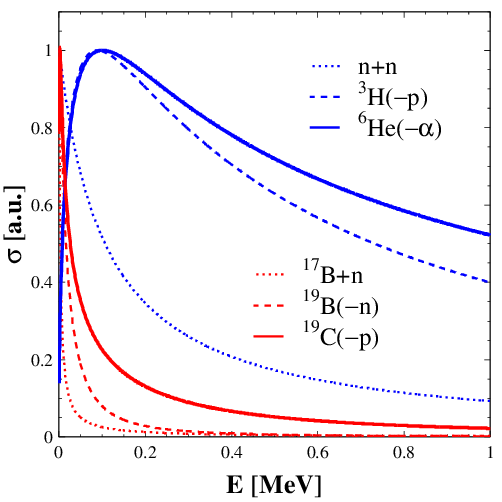,width=\columnwidth}
 \caption{Left: the phase shift $\delta_0$ of the $n$-$n$ and $^{17}$B-$n$ systems as a function of the relative energy between the particles, for the given values of $a_s$ and $r_e$ (in fm) using Eq.~(\ref{e:kcot}). Right: with those $(a_s,r_e)$ values, relative-energy spectrum of the $n$-$n$ and $^{17}$B-$n$ systems for hypothetical scattering experiments (dotted lines) using Eq.~(\ref{e:scat}), and for fast few-nucleon removal from different beams (dashed and solid lines) using Eq.~(\ref{e:AS1}). The energy spectra have been normalized to 1 at their maximum.} \label{f:obs}
\end{figure*}

 A low-energy neutron approaching a nucleus at $\ell=0$ is not subject to Coulomb or centrifugal repulsions and does not experience any potential barrier. Moreover, in the case of weakly-bound neutron-rich nuclei, the valence neutron wave function tunnels out of the potential and has a significant part in the region where it vanishes. It is thus reasonable to assume for the whole initial wave function its asymptotic form:
\begin{equation}
 \psi_i(r)\ \approx\ e^{-\alpha r}/r \label{e:phiSn}
\end{equation}
 with $\alpha=\sqrt{2\mu_i S_n}$, $\mu_i$ the reduced mass of the neutron and the rest of the initial system, and $S_n$ the neutron separation energy in that system. Inserting Eq.~(\ref{e:phiSn}) in Eq.~(\ref{e:sigmaO}) we obtain an analytical cross-section:
 \begin{equation}
 \sigma(k)\ \approx\ \frac{k}{(\alpha^2\!+\!k^2)^2}\left(\cos\delta_0+\frac{\alpha}{k}\sin\delta_0\right)^2	\label{e:AS1}
 \end{equation}
 The energy spectrum depends thus on the initial state through $\alpha$ ($S_n$) and on the final state through $\delta_0$ ($a_s,r_e$). Therefore, if the neutron binding energy in the intial system is known, a precise determination of the shape of the relative-energy spectrum should allow for the extraction of the scattering parameters of the neutron-nucleus $s$-wave interaction in the final state.

 On the left of Fig.~\ref{f:obs} we see the physical scattering observable, the phase shift, for the $n$-$n$ system (blue lines). However, as we have noted, in these unstable systems what we measure is the relative energy between the particles. On the right of Fig.~\ref{f:obs} we display (also in blue) the $n$-$n$ relative-energy spectra that one would measure, either in scattering experiments (if they were possible) either following the fast few-nucleon removal reactions $^3$H$(-p)nn$ and $^6$He$(-\alpha)nn$.
 Since the scattering lengths we are interested in should be even larger than the $n$-$n$ one, the relative-energy spectrum between nucleus and neutron would be even more shifted towards low energies, with a concentration in the first few hundreds of keV.
 
\begin{figure*}
 \begin{center}{\Large\bf\setlength{\unitlength}{.8mm}\begin{picture}(100,100)
 \Put{50}{50}{\psfig{file=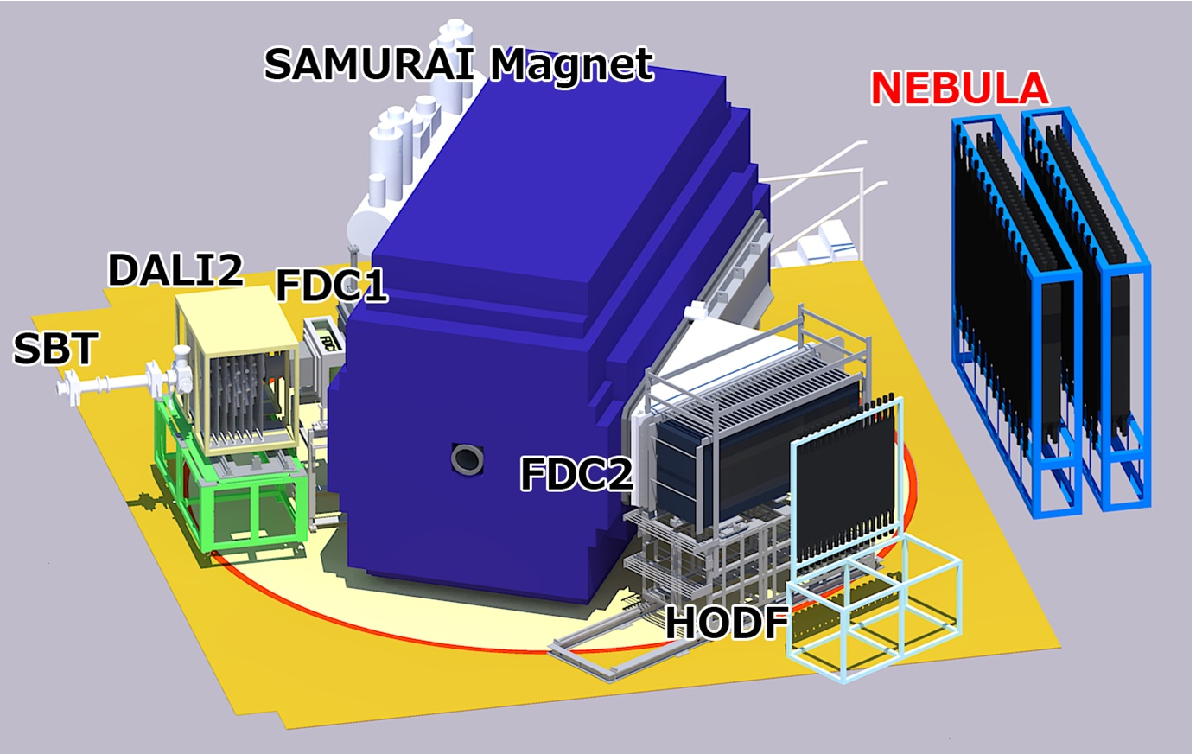,height=8cm}}
 \def\xN{102}\def\yN{68}\def\xF{83}\def\yF{37.5}\def\xT{2}\def\yT{50.8}\def\xS{-40}\def\yS{46.6}\def\neut{\point{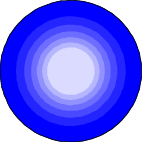}{3mm}}\def\frag{\point{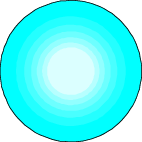}{6mm}}\def\beam{\point{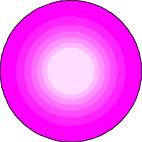}{7mm}}\linethickness{.8mm}\CadetBlue
   {\Line(\xT,\yT)(\xN,\yN)}\SkyBlue
   {\qbezier(\xT,\yT)(77,57)(\xF,\yF)}\Put{3}{52}{\fcolorbox{Black}{Yellow}{$^{18}$B}}\Lavender
   {\put(\xS,\yS){\vector(10,1){42}}}
   \Put{\xF}{\yF}{\frag} \Put{\xN}{\yN}{\neut} \Put{\xS}{\yS}{\beam}
   \Put{\xN}{61}{\fcolorbox{Black}{CadetBlue}{n}}
   \Put{\xF}{27}{\fcolorbox{Black}{SkyBlue}{$^{17}$B}}
   \Put{\xS}{36}{\fcolorbox{Black}{Lavender}{$^{19}$C}}
   \Put{\xS}{27}{\fcolorbox{Black}{Lavender}{$^{19}$B}}
   \Put{125}{24}{\Gray{\Fbox{\psfig{file=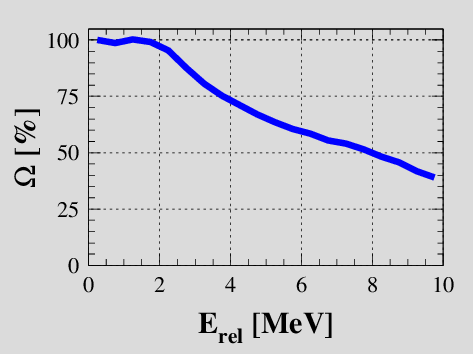,height=3.5cm}}}}
 \end{picture}}\end{center}
 \caption{Experimental setup of SAMURAI at RIKEN for the $^{18}$B campaign. Projectile beams of $^{19}$C and $^{19}$B impinge on a carbon target at $\sim230$~MeV/N; the $^{17}$B+$n$ system is produced after fast proton or neutron removal, respectively; and the $^{17}$B fragment and the neutron are detected in coincidence respectively by the FDC2/HODF and NEBULA detector arrays. The right panel shows the acceptance of the setup as a function of the relative energy between fragment and neutron.} \label{f:Day1}
\end{figure*}

\section{The Boron-17 case}

 The elementary neutron-nucleon scattering lengths are of the order of $-20$~fm, reflecting the strong attractive character of the nuclear force. Therefore, one could expect similar or higher values when adding up other nucleons to the system. However, the Pauli principle blocks the states already occupied by other neutrons and counters the pure $s$-wave attraction that the incoming neutron should feel. This delicate balance between $n$-$N$ attraction and $n$-$n$ Pauli repulsion leads, with increasing mass number, 
 to neutron-nucleus scattering lengths of only a few fm (in fact the dominant $|a_s|/r_e\sim1$ regime in Fig.~\ref{f:a_R}), even oscillating between net attraction and repulsion \cite{Jaume2020}.
 
 In a pioneering search for resonances in the $^{18}$B system at MSU, $^{17}$B$+n$ coincidences were measured following proton removal from a $^{19}$C beam at 62~MeV/N \cite{Spyrou2010}. The observation of a sharp increase of the cross-section towards zero energy was attributed to a strongly resonant $s$-wave, although experimental resolution and acceptance limitations did not allow to determine a precise value of the associated scattering parameters. 
 The effective range was not even considered and, although the data were compatible with a scattering length of $-100$~fm, only an upper limit $a_s<-50$~fm was proposed \cite{Spyrou2010}.
 
 This upper limit opens the possibility of a nuclear system actually being in the very narrow divergence windows of Fig.~\ref{f:a_R}. Moreover, $^{19}$B is a well-known two-neutron halo nucleus, with a characteristic $^{17}$B-$n$-$n$ Borromean three-body structure \cite{Cook2020}. If the $^{17}$B-$n$ scattering length were large enough, of hundreds (or even thousands) of fm, $^{19}$B states might exist in the form of Efimov trimers \cite{Emiko2019}.
 But would such a large scattering length be measurable? And what about the effective range? Even if the latter should be of the order of the nuclear radius, both length scales are needed in order to conclude about the Efimov character of a three-body system.

 Since the specific values of the $^{17}$B-$n$ scattering parameters are unknown to date, on the left of Fig.~\ref{f:obs} we draw (red lines) the phase shift corresponding to $a_s=-100$~fm, the preferred value in Ref.~\cite{Spyrou2010}, and $r_e=3$~fm, the value of the rms matter radius of $^{17}$B \cite{Rrms}.
 The larger $a_s$ (about $a_{nn}\times5$) leads to an even sharper increase of the phase shift towards $\pi/2$ at zero energy. Moreover, the effect of the range $r_e$ becomes more important, and starts at lower energies. In fact, for very large values of $a_s$ the first term of Eq.~(\ref{e:kcot}) becomes negligeable, and as a consequence the second term becomes very rapidly significant.
 Clearly, in the $^{17}$B-$n$ case the effective range must be considered in the description of the final state of the reaction. 

 On the right of Fig.~\ref{f:obs} we see (also in red) the quantity one should measure, the relative energy between $^{17}$B and the neutron. The dotted line is purely hypothetical, since such scattering experiment cannot be undertaken, but it can be compared to the dotted blue line, the analog hypothetical example for the $n$-$n$ system. The very similar lineshape obtained from Eq.~(\ref{e:scat}) is in the $^{17}$B-$n$ case about one order of magnitude lower in energy.
 Concerning the two simplest nucleon removal reactions leading to $^{17}$B+$n$ in the final state, $^{19}$C$(-p)$ and $^{19}$B$(-n)$, we see how for the same final state, and thus the same scattering parameters, the initial binding of the neutron plays a significant role in the lineshape of the cross-section.
 
\section{Boron-18 campaign at RIKEN}
 
 In order to determine the $^{17}$B-$n$ scattering parameters, an experimental campaign was undertaken at RIKEN using several beams around $^{19}$C, all leading to the same $^{17}$B$+n$ final state \cite{PhD}.
 The setup is shown in Fig.~\ref{f:Day1} (for details see for example Refs.~\cite{Cook2020,Kondo2016,Sylvain}). Beams of neutron-rich boron, carbon and nitrogen isotopes were sent onto a carbon target at an average energy of 230~MeV/N. The fast removal of nucleon(s) in the reaction was tagged by the detection of the beam-velocity fragment and neutron(s) in the forward direction. The fragment was deflected by the SAMURAI magnet, its trajectory determined by two drift chambers before and after the magnet (FDC1~and~2), and its time of flight and charge measured by a plastic hodoscope. The neutrons were detected in the NEBULA array, and their energy deduced from their time of flight. Eventual $\gamma$-ray decays in flight of the fragment were detected by the DALI2 array around the target.
 
 The previous attempt to measure the $^{17}$B-$n$ scattering parameters \cite{Spyrou2010} was hindered by a double bias, by trying to measure a very narrow energy spectrum focused at low energies (red lines on the right of Fig.~\ref{f:obs}) with a setup of moderate resolution and an acceptance restricted to those low energies.
 The RIKEN setup was designed to address those two aspects. In Fig.~\ref{f:Day1} we show the core-$n$ energy acceptance, flat in the first 2~MeV and then smoothly decreasing but significant up to 10~MeV.
 The resolution was FWHM\,$\sim0.4\sqrt{E}$~MeV, well suited for the detection of very low-energy fragment-neutron coincidences \cite{Kondo2016}.
 Therefore, any narrow low-energy lineshape should be measurable without significant distortion, allowing for the fit of the spectrum with the scattering parameters.
 
 As noted, the energy spectrum in Eq.~(\ref{e:AS1}) has only three free parameters, $\sigma(E\,|\,S_n,a_s,r_e)$.
 The initial neutron binding energy is well-known for most of the beams available \cite{AME2020}, and covers a relatively wide range.
 Among all of them, the less weakly bound is $^{21}$N ($S_n=4.61$~MeV), followed by $^{20}$C, $^{20}$N and $^{22}$N ($S_n=2.98$, $2.16$ and $1.54$~MeV), and finally the two more weakly bound cases illustrated in Fig.~\ref{f:obs}, $^{19}$C ($S_n=0.58$~MeV) and $^{19}$B ($S_n=0.09$~MeV).
 This range of different few-nucleon removal channels should therefore provide a sensitive test of the formalism, that would then lead to a more confident extraction of the scattering parameters.

 The preliminary analyses of the $^{17}$B-$n$ spectra \cite{PhD} show indeed a sensitivity to the three parameters of Eq.~(\ref{e:AS1}). Effective ranges of the order of a few fm and scattering lengths of the order of hundreds of fm lead to the best description of the different channels.
 Moreover, the use of Eq.~(\ref{e:AS1}) seems to be validated by the trend observed for the different binding energies.
 Interestingly, the $^{19}$B$(-n)$ channel exhibits a lineshape very similar to the $^{19}$C$(-p)$ one, despite the difference in binding energy (90 and 580~keV respectively).  
 However, the present error for $^{19}$B ($S_n=90\pm^{560}_{90}$~keV \cite{AME2020}) is larger than that of the other nuclei. If the actual energy was close to its upper limit (roughly similar to that of $^{19}$C), this could explain the similarity of their lineshapes, and would in fact be consistent with results on the electric dipole strength of $^{19}$B \cite{Cook2020}.

\section{Summary and perspectives}

 The Efimov effect emerged as an intriguing universal prediction in the 1970s \cite{Efimov1970}. Strongly resonant interactions between two bodies, barely unable to bind them, could lead under certain circumstances to an infinite series of three-body bound states. These circumstances, however, were possible but very unlikely to find in nuclear physics, and the experimental confirmation of the effect remained elusive for 35~years. The effect being universal, it was finally confirmed in atomic physics, in several mixtures of ultra-cold atom gases, thanks to the tuning of the interaction between atoms through magnetic fields.
 
 Contrary to atomic systems, two-body interactions in nuclear systems cannot be externally tuned, and therefore one must rely in the unlikely finding of a neutron-nucleus system with an extremely large scattering length, well beyond the nuclear scale. 
 In order to explore the most exotic neutron-nucleus combinations, scattering experiments must be replaced by high-energy few-nucleon removal reactions, able to produce any neutron-nucleus system in the final state at low relative energies. The high energy of the reaction allows the use of the sudden approximation, in which the cross-section is calculated as the overlap integral between the wave functions of the initial bound state and of the final scattering one.
 
 An experimental campaign aiming to explore the most promising candidate, the $^{17}$B-$n$ system \cite{Spyrou2010}, has been undertaken at RIKEN \cite{PhD}.
 The high resolution and acceptance of the setup, together with the variety of initial states used, should allow us to measure for the first time the value of $(a_s,r_e)$ for the $^{17}$B-$n$ system, with preliminary results that point towards an unprecedented ratio of $|a_s|/r_e\sim100$ in nuclear physics \cite{PhD}. According to Efimov's prediction, such a ratio would lay within the range of existence of one and two trimers (Fig.~\ref{f:Ooo}), but it remains to be confirmed whether there are states in $^{19}$B that would correspond to Efimov's prediction \cite{Emiko2019}.
 
 The unambiguous determination of $(a_s,r_e)$ from the few MeV of the $^{17}$B-$n$ energy spectrum could also open the way to the exploration of the third term of Eq.~(\ref{e:kcot}), associated with the shape of the potential, at slightly higher relative energies. Moreover, the suggested sensitivity of the neutron-nucleus energy spectrum to the binding of the neutron in the initial state might be used backwards to constrain this binding energy for cases in which it is poorly known, such as $^{19}$B \cite{AME2020}.
 The analysis of the different reaction channels is in progress \cite{PhD}.


\begin{thebibliography}{}
\bibitem{Hammer2006} E.~Braaten and H.W.~Hammer, {\it Universality in few-body systems with large scattering length}, Phys.\ Rep.\ {\bf428}, 259 (2006).
\bibitem{Efimov1970} V.~Efimov, {\it Energy levels arising from resonant two-body forces in a three-body system}, Phys.\ Lett.\ {\bf33B}, 563 (1970).
\bibitem{Scattering} W.F.~Hornyak, {\it Nuclear structure} (Academic Press, New York, 1975), pp.~63-73.
\bibitem{Pascal2017} P.~Naidon and S.~Endo, {\it Efimov physics: a review}, Rep.\ Prog.\ Phys.\ {\bf80}, 056001 (2017).
\bibitem{Efimov1973} V.~Efimov, {\it Energy levels of three resonantly interacting particles}, Nuc.\ Phys.\ {\bf A210}, 157 (1973).
\bibitem{Endo2025} S.~Endo \etal, {\it Three-body forces and Efimov physics in nuclei and atoms}, Eur.\ Phys.\ J.\ A {\bf61}, 9 (2025).
\bibitem{FMM2021} F.M.~Marqu\'es and J.~Carbonell, {\it The quest for light multineutron systems}, Eur.\ Phys.\ J.\ A {\bf57}, 105 (2021).
\bibitem{Feshbach} C.~Chin \etal, {\it Feshbach resonances in ultracold gases}, Rev.\ Mod.\ Phys.\ {\bf82}, 1225 (2010).
\bibitem{Efimov_CsCsCs} T.~Kraemer \etal, {\it Evidence for Efimov quantum states in an ultracold gas of caesium atoms}, Nature {\bf440}, 315 (2006).
\bibitem{Efimov_x2} B.~Huang \etal, {\it Observation of the second triatomic resonance in Efimov's scenario}, Phys.\ Rev.\ Lett.\ {\bf112}, 190401 (2014).
\bibitem{Efimov_RbKK} G.~Barontini \etal, {\it Observation of heteronuclear atomic Efimov resonances}, Phys.\ Rev.\ Lett.\ {\bf103}, 043201 (2009).
\bibitem{Efimov_x3a} R.~Pires \etal, {\it Observation of Efimov resonances in a mixture with extreme mass imbalance}, Phys.\ Rev.\ Lett.\ {\bf112}, 250404 (2014).
\bibitem{Efimov_x3b} S.K. Tung \etal, {\it Geometric scaling of Efimov states in a $^6$Li-$^{133}$Cs mixture}, Phys.\ Rev.\ Lett.\ {\bf113}, 113 (2014).
\bibitem{Jaume2020} J.~Carbonell \etal, {\it Low-energy neutron scattering on light nuclei and $^{19}$B as a $^{17}$B-n-n three-body system in the unitary limit}, SciPost Phys.\ Proc.\ {\bf3}, 008 (2020). 
\bibitem{Spyrou2010} A.~Spyrou \etal, {\it First evidence for a virtual $^{18}$B ground state}, Phys.\ Lett.\ B {\bf683}, 129 (2010).
\bibitem{Cook2020} K.J.~Cook \etal, {\it Halo structure of the neutron-dripline nucleus $^{19}$B}, Phys.\ Rev.\ Lett.\ {\bf124}, 212503 (2020).
\bibitem{Emiko2019} E.~Hiyama \etal, {\it Modeling $^{19}$B as a $^{17}$B-n-n three-body system in the unitary limit}, Phys.\ Rev.\ C {\bf100}, 011603R (2019).
\bibitem{Rrms} T.~Suzuki \etal, {\it Nuclear radii of $^{17,19}$B and $^{14}$Be}, Nucl.\ Phys.\ A {\bf658}, 313 (1999).
\bibitem{PhD} E.~Oliveira, {\it The structure of the most neutron-rich Boron isotopes}, PhD thesis (2025), Normandie Universit\'e, {https://theses.hal.science/tel-05455987v1}. 
\bibitem{Kondo2016} Y.~Kondo \etal, {\it Nucleus $^{26}$O: a barely unbound system beyond the drip line}, Phys.\ Rev.\ Lett.\ {\bf116}, 102503 (2016).
\bibitem{Sylvain} S.~Leblond \etal, {\it First observation of $^{20}$B and $^{21}$B}, Phys.\ Rev.\ Lett.\ {\bf121}, 262502 (2018).
\bibitem{AME2020} M.~Wang \etal, {\it The AME 2020 atomic mass evaluation}, Chinese Phys.\ C {\bf45}, 030003 (2021).
\end{thebibliography}
\end{document}